# $MgB_2$ synthesis in closed volume and pinning potential.

Anan'ev S. P., Grinenko V. A., Keilin V. A., Krasnoperov E. P., Levit P. A., Stoliarov V. A.

RRC Kurchatov Institute, 123182 Moscow.

*Annotation*

*Superconducting $MgB_2$ was synthesized in a stainless steel tube which compressed by pulsed magnetic field. The temperature dependence of diamagnetic moment m(T) was investigated and the pinning potential was determined.*

PACS N 74.25.Qt.

The magnesium diboride ($MgB_2$) which supercoductivity was discovered last year [1] has been synthesized by different methods [2]. The most popular method of $MgB_2$ fabrication is a high temperature (T≈ 900C) solid state reaction in inert atmosphere [2,3]. In this method Mg and B powders are pressed into pellet wrapped with Ta or stainless steel foil. The pellet are heated up to temperature 900-930° C in Ar atmosphere. The main deficiency of this method is a Mg loss from pellet due to high pressure of magnesium vapours at synthesis temperature. From pellet of volume 0.05 cm$^3$ Mg loses can excess 10%. To get after reaction the nominal composition ratio (Mg to B ratio = 1:2) the extra magnesium must be added. Obviously in order to compensate the loss of Mg at high temperature and to retain the initial ratio of powders the reaction should be carried out in vessel which walls are not permeable for Mg and B. One of such methods is known as "powders in tube" synthesis [4,5]. But powder must be compressed homogeneus as high as possible. In our work we developed abovemention method and pressed powders in tube by pulsed magnetic field. Powders of boron and magnesium (with Mg abundance ≤ 1%) were jam-packed by "rod and hammer" in stainless steel tube (outer diameter 4 mm, wall thickness 0.3 mm) encapsulated in a copper tube 1 mm thick. To make a tube hermetic its ends were closed by iron plugs and welded with pulsed "cold welding" [6]. Then a pulse of magnetic field was applied (the amplitude 50T and duration few micro second) and the copper tube and powders were pressed.

Synthesis of $MgB_2$ was carried out at temperature 900° C during 1-2 hours. By weighing we found that mass of sample reduced less than 0.4%. X-ray diffraction analysis showed that the content of $MgB_2$ phase was higher than 90%. Superconducting (resistive) transition measured by the standard four- probe technique is shown in the right top part of Fig.1. For magnetic measurements we used a cylindrical sample of approximately 8mm length, which was welded to the copper finger of the cryocooler. Ge thermometer was set near the sample. Experimental procedure was as follows. The sample was cooled at zero magnetic field down to the temperature lower than the critical one (ZFC), then steady magnetic field (H ≈ 0.2 T) was applied. Magnetic field $H_t$ at the sample end was measured by Hall probe. This field is equal to a half of induction in the centre of a long cylinder. The Hall probe constant changes very small in the temperature range 10-45K, so from measured field the magnetic moment $4\pi\ m(T) = \alpha\ H_t - H$ can be easily determined, where $\alpha$ is demagnetization factor. The temperature dependencies of the measured field $H_t(T)$ in the heating process from a few temperatures (ZFC) are shown in Fig. 1. It is easy to see that the diamagnetic moment values are practically independenton on the way how a proper sample state was reached: as a result of heating from lower temperature or magnetization at this temperature. The curves $H_t$ (T) are also the same at heating rates from 0.001K/s to 0.02K/s. Thus, we suppose the sample was in a temperature equilibrium. Some lower diamagnetic moment at magnetization at fixed T is caused by heat generation due to flux flow. This heat is negligible at lower field rates.

The critical current density $J_c$ can be correctly deduced from these measurements if during a measurements the creep of the magnetic flux is negligible [7]. Usually magnetic moment reduces logarithmically with time

$$m(t)/m(0) = 1 - S\ \ln(1 + t/\tau) \quad (1)$$

where S is a normalized relaxation rate (creep velocity) and $\tau$ is a characteristic time. Figure 2 shows S(T) for 1000s measurement time. At lower temperatures (T<30K) relaxation rate is small and does not exceed 0.003. Near the critical temperatures S increases considerably. These data coincide with results of others authors (see. f.e. [8]). Figure 3 shows the time evolution of induction at fixed temperature and changes of $H_t$ when T is increased. At any time temperature is fixed at value T=35.02K and during 2000s the moment decease was observed. Then the temperature was increased, but the moment practically did not change, till T reached a value corresponding to the curve m(T). From this data we can conclude

- m(T) curve is equilibrium,

- magnetic field distribution into the sample remains the same both during creep process and heating.

As at fixed temperature we can not observe any flux flow during 100s we consider the critical current (at least in this time interval) is determined by condition that the average potential barrier due to Lorentz force is equal to zero [7]. According to the critical state (Bean model) the magnetic field inside the cylindrical superconductor changes linearly with radius and current density is constant and equal to the critical value $(J=J_c)$. In this case m(T) dependence is the same as $J_c$ (T). With heating the critical field reduces and the diamagnetic moment goes to zero at $J(\bullet_c)=0$. In this measurement method the critical temperature $\bullet_c$ characterizes not the transition to the superconducting state but an appearance of resistive losses in superconducting phase. It is evident that $\bullet_c$ in the resistive method is higher than in the magnetic one.

Using a linear dependence between moment and critical current it is easy to find temperature dependence of a pinning potential U(T). It is well-known interpolation expression for U in normalized coordinate [8-10]

$$U(T) = \frac{U_0 (1-(\frac{T}{T_x})^2)^2}{(H/H_{c2})^n} \quad (2)$$

where $T_x$ is a critical temperature in magnetic field H. Supposing the critical field $H_{c2}$ has a parabolic temperature dependence the equation (2) was used to estimate value of n. It was found that the best approximation corresponds to n=2, $T_x$ =35.1K and $T_c$ =37K. This curve is shown in Fig.1. For low magnetic fields this function agrees with $J_c$ (T) reported by other authors [11]. It can be noticed that this function differs from the Kim-Anderson model function $U \propto H^{-1}$ (n=1), that describes low temperature superconductors such as Nb alloys [10].

Function U(T) have a negative curvature $(\partial^2 U(T)/\partial \cdot^2 < 0)$. It points out that the grains in the sample appear to be well linked [12]. On the other hand it was found that a considerable this resistance between superconducting $MgB_2$ core and stainless steel tube exists. The origin of resistance can be result of volume reduction during reaction or content distortion in the interface between $MgB_2$ and tube walls. An experiment to study the influence of a pressure on the value of intergranular conductivity was carried out. The commercial powder $MgB_2$ was pressed by 5 kbar into pellet. In Fig.1 induction within the pellet vs. temperature is presented. It can be seen that the magnetic moment in the pressed powder is much less than in $MgB_2$ synthesized in the tube, and its temperature dependence is almost linear $m(T) \propto (T-T_c)$. A similar dependence is observed in granular high-temperature superconductors [13,14]. It is well known that if a dielectric layer separates two superconductors then highest non dissipative Josephson current increases linearly when the temperature reduces [15]. It is possible that in pressed powder in spite of $MgB_2$ metal conductivity of grains their surfaces are distorted and present a barrier for electron penetration.

Summarizing, pulsed high magnetic field was successfully used to fabricate $MgB_2$ in tube. This method provides good contacts between the $MgB_2$ superconducting grains. Obtained results could be very useful for practical applications.

We acknowledge Dr. A. A. Bush for commercial $MgB_2$ powder supplying.

MgB2 synthesis in closed volume and pinning potential.

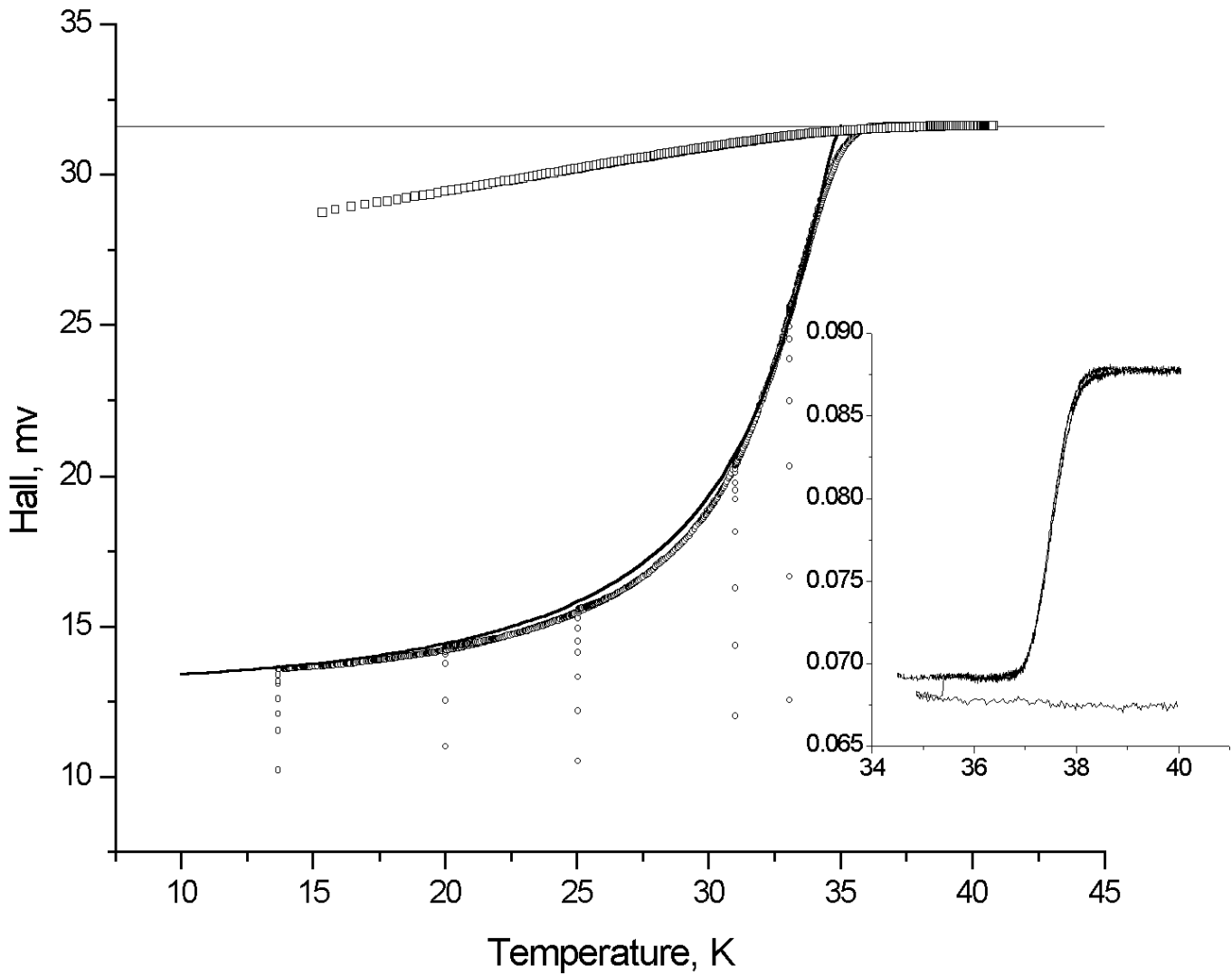

Fig.1. Temperature dependence of $MgB_2$ sample induction in field $\approx 0.2$ T.

a) o - $MgB_2$ sample synthesized after high magnetic field pressing

b)  - pressed commercial $MgB_2$ powder.

Solid curve is approximation function (2) with n=2.07, $T_x$ =35.1K and $T_c$ =37K. In right top side of figure a resistive superconducting transition is shown.

MgB2 synthesis in closed volume and pinning potential.

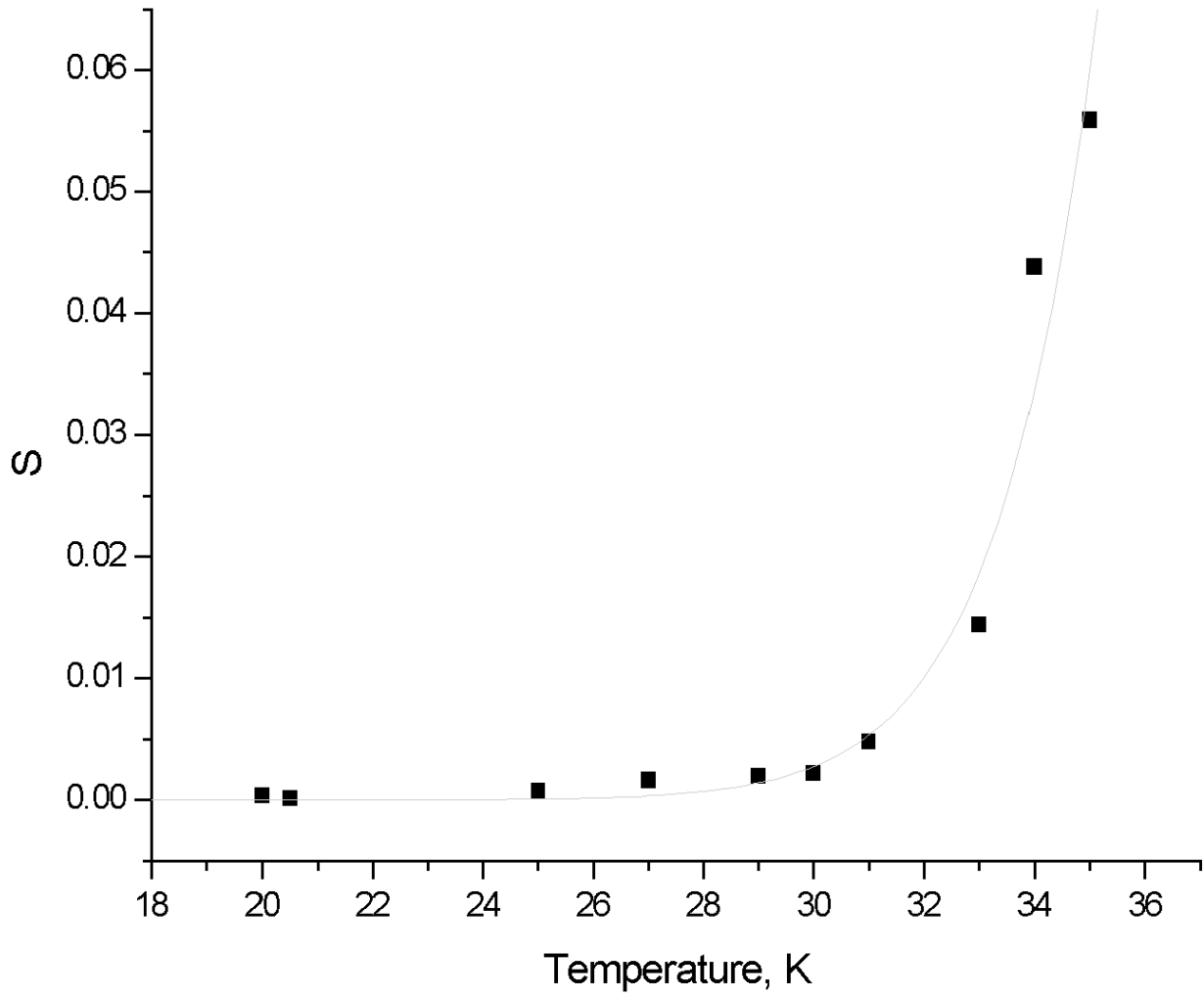

Fig.2. Long- term relaxation rate S vs. temperature.

MgB2 synthesis in closed volume and pinning potential.

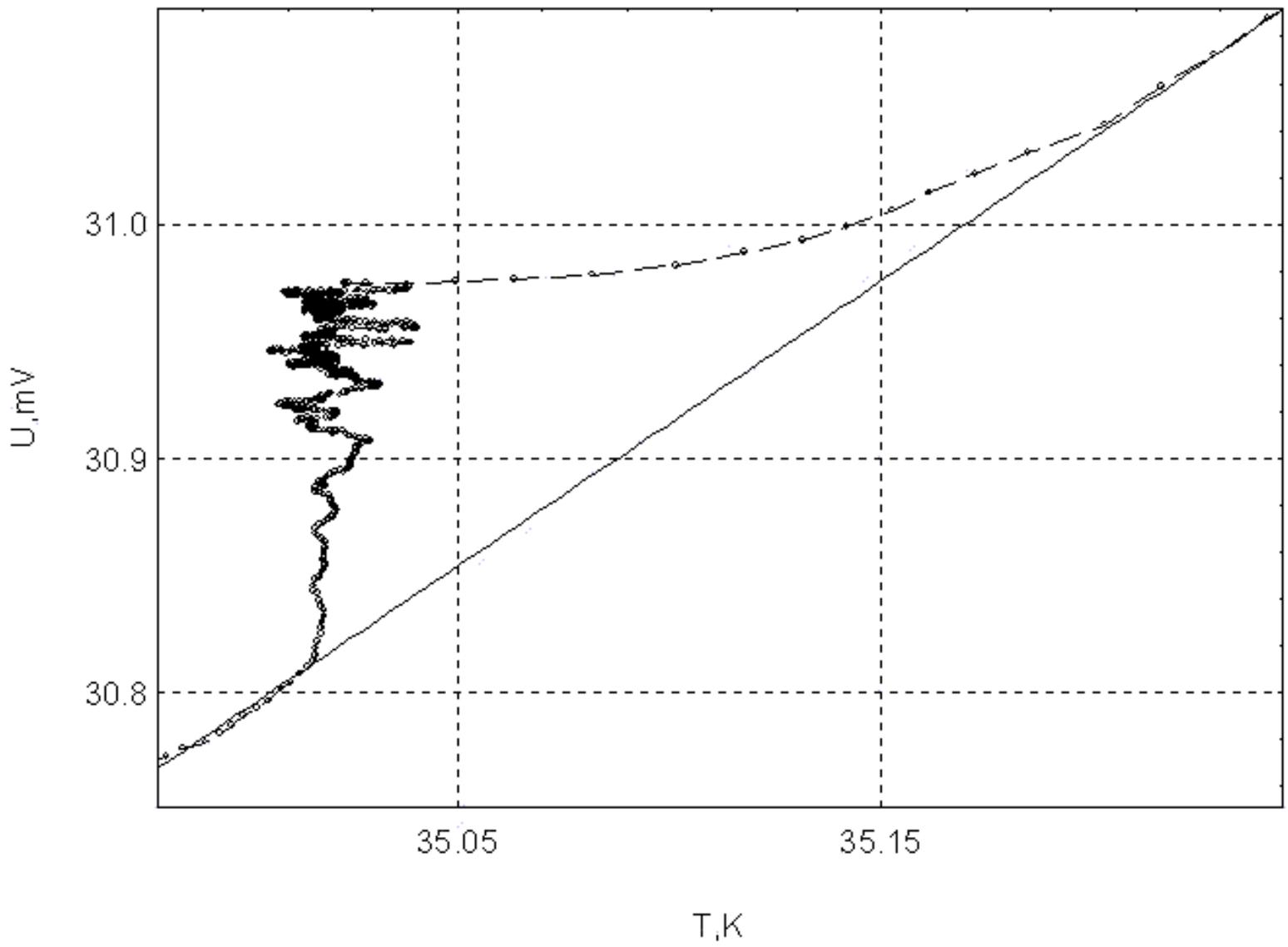

Fig.3. Behaviour of induction B(T) after creep at a fixed temperature during $10^4$ s.